# A Comprehensive Review on Digital Image Watermarking

Shweta Wadhera[a,b*], Deepa Kamra[c], Ankit Rajpal[d], Aruna Jain[e] and Vishal Jain[f]

[a] *Department of Computer Science, Deen Dayal Upadhyaya College, University of Delhi, New Delhi, India*
[b] *Research Scholar, Sharda University, Greater Noida, India*
[c] *Department of Management Studies, Deen Dayal Upadhyaya College, University of Delhi, New Delhi, India*
[d] *Department of Computer Science, University of Delhi, New Delhi, India*
[e] *Department of Computer Science, Bharati College, University of Delhi, New Delhi, India*
[f] *School of Engineering and Technology, Department of Computer Science and Engineering, Sharda University Greater Noida, India*

**Abstract**
The advent of the Internet led to the easy availability of digital data like images, audio, and video. Easy access to multimedia gives rise to the issues such as content authentication, security, copyright protection, and ownership identification. Here, we discuss the concept of digital image watermarking with a focus on the technique used in image watermark embedding and extraction of the watermark. The detailed classification along with the basic characteristics, namely visual imperceptibility, robustness, capacity, security of digital watermarking is also presented in this work. Further, we have also discussed the recent application areas of digital watermarking such as healthcare, remote education, electronic voting systems, and the military. The robustness is evaluated by examining the effect of image processing attacks on the signed content and the watermark recoverability. The authors believe that the comprehensive survey presented in this paper will help the new researchers to gather knowledge in this domain. Further, the comparative analysis can enkindle ideas to improve upon the already mentioned techniques.

**Keywords 1**
Digital Watermarking, Digital images, review, visual imperceptibility, robustness, copyright protection

## 1. Introduction

In the present era, the secret to the success of an organization is all about the information it can gather. Also, of concern is how effectively it can stop others from accessing the information generated through its operations and processes. The popularity and availability of the Internet, easy access to digital storage devices, have made the creation, replication, and distribution of digital media hassle-free. This has led to the strong need of developing methods for preventing copyright breaches [1, 2].

The technique of digital watermarking is being applied widely to the situations where an organization wants to deter the data from spilling into the public domain. It is extremely crucial in cases where the company is in direct fiduciary relationship with its customers and must protect their information as well [1].

Digital Watermarking is a method to provide protection from any tampering or alteration [3, 4, 5]. It provides security and authentication to digital content. The digital watermarking process involves the insertion of signal, information into the original media content. The inserted information is then uncovered and extracted to report the actual owner of the digital media.

Watermarking involves embedding data called watermark or digital signature or label into the digital media. This watermark can be extracted for revealing the authenticity of the media object [6]. As an example of a watermark, we can imagine a visible "seal" over an image. A digital watermarking algorithm has three basic parts: -

1. Watermark
2. The Encoding algorithm
3. The decoding algorithm [1, 7, 8].

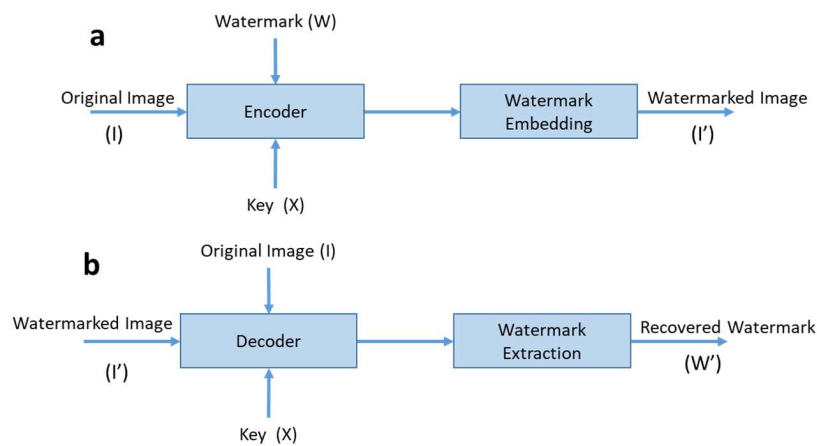

**Figure 1:** The block diagram of Digital Watermarking Concept

This technique may be used for copyright safeguarding and embedding fingerprints, placing authentication for data integrity checks, and confidential communication. Another important application of watermarking techniques includes tracing illegal users with an objective that the owner can approach the regulatory authorities. It can be useful for ensuring that data pertaining to people who buy and sell digital media, kept in record for each transaction. Further track of this data can be kept for controlling copyright breaches. In fact, strict measures must be implemented for this unlawful distribution of digital content.

The paper is consolidated as follows: section 2 presents the classification of digital watermarking based on various characteristics. Section 3 describes the prominent features of digital watermarking. Section 4 covers the recent areas of application for watermarking. Section 5 presents the various types of attacks. In section 6 we analyze the highlights and results of related work by various authors and section 7 presents the conclusions and future scope in the direction of image watermarking.

## 2. Classification of Digital Watermarking

The section describes the classification of digital watermarking based on various criteria such as robustness, perceptibility, domain, and detection process, multimedia. Further we also briefly present the different popular watermarking techniques.

## 2.1. Based on Characteristics/Robustness

**Robust:** Robust watermarking is preferred when copyright information is required to be inserted. Robustness is indicated if the embedded watermark even after some attack is not damaged. It can withstand several attacks. It is seen that for copyright protection, a robust watermark is advantageous.

**Fragile:** One can easily find out from the status of the watermark if the data has been altered. For integrity protection, this watermark is preferred.

**Semi-fragile:** Some extent of change is tolerated by a semi-fragile watermark [2, 3, 6].

## 2.2. Based on Perceptibility

**Perceptible:** A watermark that is visible is known as perceptible.

**Imperceptible**: Incase the watermark is invisible then it is known as imperceptible watermark. In this, information which is embedded into the image is not visible. In such cases one can prove the ownership of your image with the help of imperceptible watermark [2, 26].

## 2.3. Based on Domain

**Frequency domain:** First, the transforming of the image to the frequency domain is carried out. In this type of watermarking different transformation techniques are applied such as DCT, DFT, and DWT [27].

**Spatial Domain:** Watermarking in this domain moderately changes the value of pixels, of arbitrarily selected portions of images; the watermark is inserted in the host image. No conversion or transformation is applied in the spatial domain. LSB, Patchwork method, SSM Modulation are some of the popular spatial domain-based techniques [2, 3].

Generally, the watermarking carried out in the frequency domain is more robust as compared to the one carried out in the spatial domain.

### 2.3.1. Spatial Domain

**Least Significant bit (LSB) method**

In this method, the watermark is inserted in the least significant bit (LSB of image pixels. Ideally, either of two ways is used for embedding. In one approach, the LSB of an image is substituted with a pseudo-noise (PN) sequence, while another approach adds this PN sequence to the LSB. LSB technique provides ease of use but compromises on the robustness parameter against attacks.

**Patchwork Method**

Patchwork method randomly picks n pairs of image points (x, y). The data in the x region is lightened and data in the y region is darkened. However, this technique can withstand a series of attacks, but it lacks in terms of capacity.

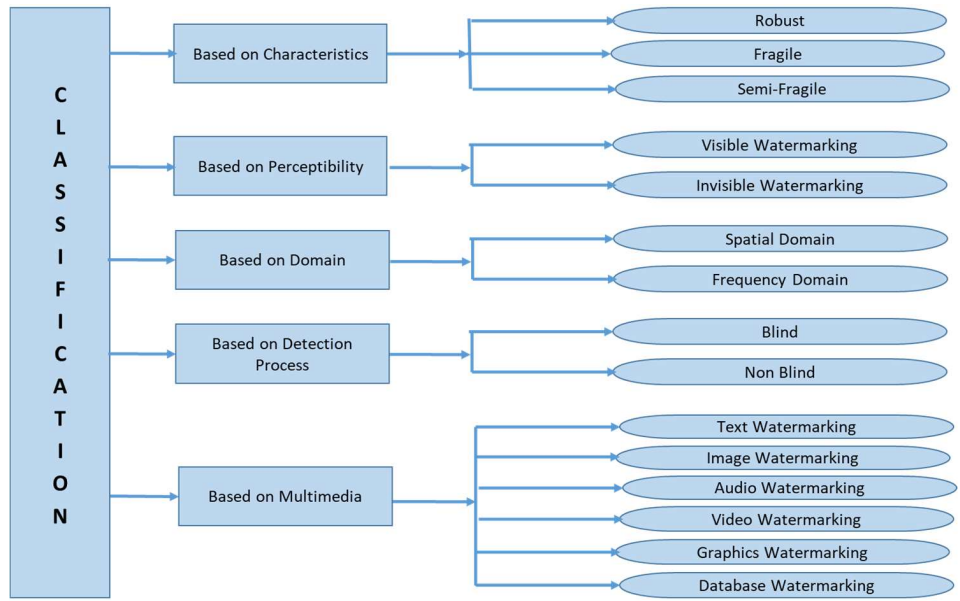

**Figure 2**: Classification of Digital watermarking

## 2.3.1. Frequency Domain

**Discrete cosine transforms (DCT)**

This method employs a technique of subdividing an image signal into non-overlapping blocks of 8 × 8 size. Further, block-wise DCT is performed, and thereafter the choice of coefficients to be watermarked is carried out. Finally, an inverse DCT is applied on each 8 × 8 block to obtain the signed image.

**Discrete wavelets transform (DWT)**

In this method, the image is subjected to a sequence of low-pass and high-pass filters. An image is decomposed into four equal sub-bands where each sub-band comprises low frequency (LL), horizontal features (LH), vertical features (HL), and diagonal features (HH). It is a preferred algorithm as it provides a robust and secure watermarking method [2, 3, 27].

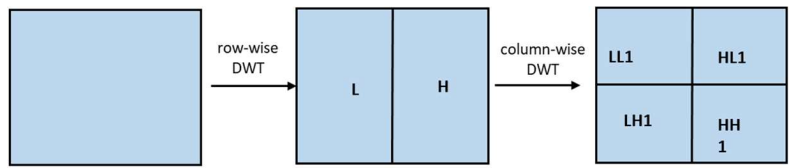

(a) First Level of decomposition

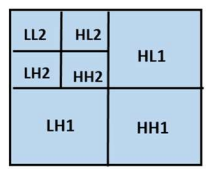

(b) Second Level of decomposition

**Figure 3:** Three level decomposition in 2D DWT

## 2.4. Based on Detection process

- **Blind:** Those watermarking processes fall in this category wherein the removal of the embedded information requires only the watermarked image. Its applications can be copyright protection, e-voting, etc. [9].

- **Non-Blind:** In this type of watermarking, the process copies along with the text data, the host image, and the inserted information for the retrieval of the watermark. Its application is seen in copyright protection [9].

## 2.5. Based on Multimedia

- **Text Watermarking:** It consists of components like words, punctuations, sentences, etc. Transformation is done on one of these components and is embedded as a watermark [3].

- **Image Watermarking:** Large-size images are there which has to be watermarked. In the case of images, we require robust watermarks, which should be imperceptible [3].

- **Video watermarking:** In this case, it is difficult to get imperceptible watermark.

- **Graphic Watermarking:** In 2D or 3D digital graphics, a watermark is embedded. It provides copyright protection.

## 3. Basic Characteristics of Digital Watermarking

Mentioned here are various features of Digital Watermarking.

- **Robustness:** The robustness feature indicates that the digital watermark can resist various processing operations and attacks. Then it is considered to be robust [14].

- **Imperceptibility:** The imperceptibility feature indicates that a digital watermark should not be seen by the human eye. One should not be able to see the embedded watermark. It can only be identified by specialized procedures. The watermark should be such that the viewer should not be able to see it and the process of embedding a watermark should be such that the quality of the content is maintained [14].

- **Security:** The security feature indicates that irrespective of targeted attacks, the inserted digital watermark cannot be removed. Watermark security describes that altering or removing a watermark without any deterioration to the host signal should be arduous. Watermarking security can be explained as a way to provide secrecy, ownership, and protection of data [2].

- **Capacity:** The amount of information embedded in a watermarked image also known as data payload [3].

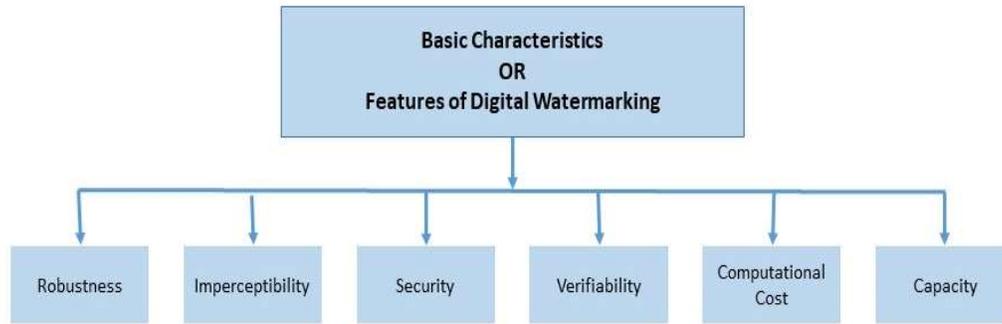

**Figure 4:** Basic Characteristics

- **Verifiability:** Through the watermark, we should be able to get a piece of evidence regarding the ownership of copyright-protected data. This helps in verifying the authenticity of any digital data and even the control of its unlawful copying [29].

## 4. Recent Applications of Watermarking

- **Copyright protection:** As we are aware that images can be easily circulated and are freely available over the internet. These images can be used commercially. So copyright protection of data is needed and for this Digital Watermarking is very useful. The inserted digital watermark will be used to identify the copyright owner [2, 9].

- **Fingerprint:** Fingerprinting in digital watermarking can be put up as a method for embedding some distinctiveness. The fingerprint should be difficult to alter. The information inserted is related to the customer. It is through this fingerprinting that it is revealed about those authorized customers who are involved in the circulation of copyright data by breaching the agreement [2, 27].

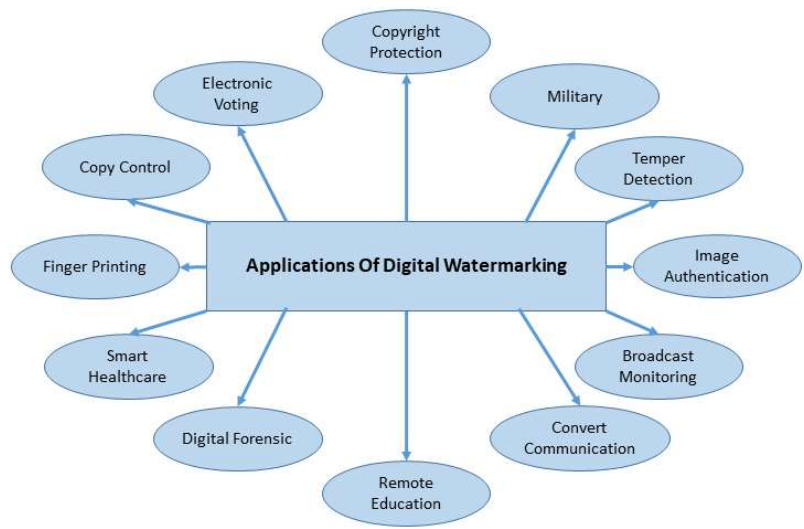

**Figure 5:** Applications of Digital Watermarking

- **Copy control:** Digital watermarking can prevent illegal duplication of digital data. Devices that do replication can detect these watermarks and report copying and thereby put a control on illegal copying [3, 26].

- **Broadcast Monitoring:** Over the years it has been seen that availability and accessibility to media content have increased exponentially. Also, the content is available through the internet. In such times, it has become important for content owners and copyright owners to know about the real distributor of content. Digital Watermarking has an important role here [3, 26].

- **Medical Application:** Embedding of patient name in the MRI-scan, CT-scan or X-ray reports can be done using visible watermarking. The treatment of the patient depends on these medical reports. So to avoid mixing of reports the technique of visible watermarking can be used [9].

- **Electronic Voting System:** The Internet has spread all over the country from big towns to small villages. Electronic voting helps to carry out elections, keeping the security aspect into consideration [9].

- **Remote Education:** Lack of teachers poses a big problem in small villages. Smart Technology needs to be adopted for distance learning. In this case watermarking plays its role in the authentic transmission of study material over the internet [9, 27].

## 5. Attacks on Watermarking

A digital Watermarking scheme is always assessed by the fact that how robust it is over attacks. Attack on any watermark is used to harm the inserted watermark or enfeeble the watermark's discovery. Hampering the protection, which a watermark provides to digital content, is the objective of any attack. Watermarking attacks can be classified as follows – Geometry attack, Protocol attack, Cryptographic attack, and removal attack [10, 11, 12].

- **Geometry Attack:** Such processing is done over the watermark image which alters the geometry of the image like rotation, cropping, etc. These can be further classified into – scaling, cropping, rotation, and translation [6].
- **Removal Attack:** This attack aims at removing the inserted data from the digital image. If it is not able to, yet they try to destroy the embedded information [29].

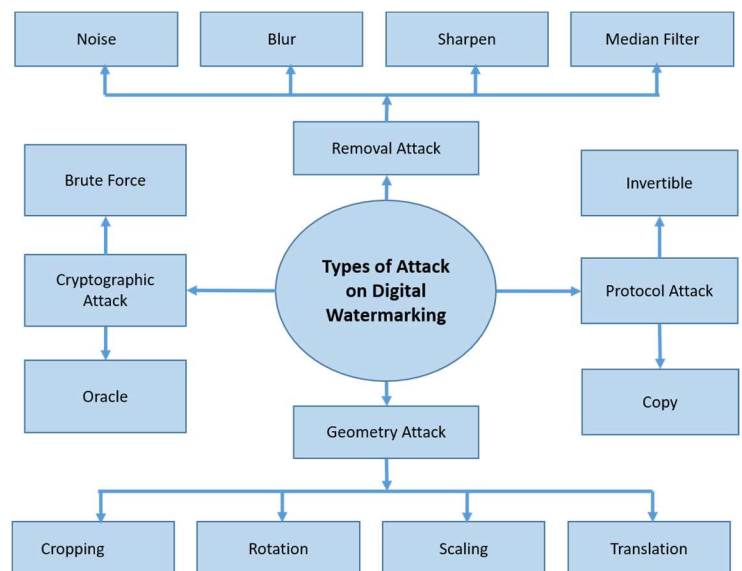

**Figure 6:** Types of Attacks on Digital Watermarking

- **Protocol Attack:** The attacks which come under this category, do not damage the embedded data. Two types of Protocol attacks are there: Invertible, Copy attack. So a watermark should be non-invertible and should not be copied. A watermark is invertible when the attacker removes his own watermark from the host data. The attacker then pretends to be the owner of the data. This shows that for copyright protection, watermarks should be non-invertible [13, 29].

- **Copy Attack:** It is also a form of Protocol attack. In this also the watermark is not destroyed. Instead, the attacker estimates the watermark from host data. It is then copied to other data [13].

- **Cryptographic Attack:** These types of attacks include those which break the security in watermarking techniques. With this, they can extract the inserted watermark data or can insert some delusive watermark. The Brute-force and Oracle attacks fall under this category [29].

In order to present an exhaustive survey on image watermarking, we created a repository of more than 50 papers on Mendeley and 35 papers were used to develop the background of digital watermarking and 19 papers were found relevant in the area of image watermarking.

Table 1 summarizes the watermarking schemes proposed by various research groups in the past few years through the comparative analysis.

**Table 1:** Comprehensive survey of recent image watermarking schemes

| Research group | Title | Technique used | Input | Visual Imperceptibility | Robustness |
|---|---|---|---|---|---|
| Abraham and Paul [6] | "An imperceptible spatial domain color image watermarking scheme" | Spatial domain Simple Image Region Detector (SIRD): Estimation of most suitable portion within the block of an image. | **Cover image** Colored image of Size: 512 x 512 x 3 pixels  **Watermark** Size: 64 x 64 pixels | PSNR = 47.6 dB  SSIM = 0.9904 | NCC = Range [0.9917 – 1]  BER= Range [0.7500-0]  **Attacks considered:**  Salt and Pepper, Poisson, Speckle, average filtering, Gaussian LPF, Sharpening, JPEG Compression, Cropping, Resizing |
| Liu et al. [7] | "Secure and robust digital image watermarking scheme using | Scrambling Watermark: **RSA Encryption**  **DWT + SVD Hybrid** | **Cover image** Colored Size: 512 x 512 pixels | PSNR Range = [38.68 -48.03] dB | NCC Range =[0.6053 - 0.9673]  **Attacks** |
| | logistic and RSA encryption" | Different embedding strength is used. | **Watermark Size:** 256 x 256 pixels | | **Considered:** Mean Filtering, Median Filtering, Gaussian Noise, Salt & Pepper Noise, Rotation, Crop, JPEG compression |
| Vaidya and Mouli [14] | "A robust semi-blind watermarking for color images based on multiple decompositions" | **DWT-CT-Schur-SVD** **Discrete Wavelet Transform (DWT):** 1-level decomposition Contour Let Transform (CT): 2-level on LL subband to obtain $D_1^2$. **Schur Transform:** $D_1^2$ is decomposed as $D_1^2 = Q_1^2 . P_1^2 . [Q_1^2]^T$ **Singular Value Decomposition (SVD):** Watermark is inserted in $P_1^2$. | **Cover image** Colored Two Watermarks of Size: 64 x 64 pixels | Watermark1 PSNR range = [27.63-36.16] dB SSIM range = [0.9709-0.9989] Watermark2 PSNR Range =[22.38- 31.59] dB | NCC=1.0 **Attacks Considered:** Salt & Pepper, Gaussian Noise, |
| Liu et al. [15] | "Digital image watermarking method based on DCT and fractal encoding" | Fractal encoding and DCT method are combined for double encryption for embedding purpose. | **Cover image Size:** 1024 x 1024 pixels **Watermark Size:** 256 x 256 pixels | PSNR Range=[41-45 ] dB | **Attacks Considered:** white noise attack, Gaussian filter attack, JPEG compression attack. |
| Savakar and Ghuli [16] | "Robust Invisible Digital Image | **Secret key:** Select the place of insertion of the | **Cover Image** Plenoptic Size: | PSNR = Range[51.68-64.54] dB | NCC= Range[0.9139-1.0] |
| | Watermarking Using Hybrid Scheme" | logo. DCT + SVD Hybrid | 512 x 512 pixels **Watermark** Binary Size: 50 x 20 pixels | SSIM= Range[0.9989-1.0000 ] | **Attacks Considered :** Gaussian Noise, Speckle noise, Salt and pepper noise, Poisson Noise, Rotation, JPEG Compression |
| Moosazadeh and Ekbatanifard [17] | "A new DCT-based robust image watermarking method using teaching-learning-Based optimization" | **Image watermarking scheme based on DCT** **Teaching-Learning-Based Optimization (TLBO):** Automatic detection of embedding parameters and suitable position for inserting the watermark. | **Cover image** Size: 512 x 512 pixels **Watermark** Size: 32 x 32 pixels | PSNR= Range[39.95-40.73] | NCC = Range [0.7871-0.9901] **Attacks considered :** Salt & pepper noise, Uniform noise, Poisson noise, Gaussian noise, Scaling, Rotation, Cropping, Sharpening, Motion Filter, Disk filter, Wiener, Median filter, Gaussian Filter, JPEG compression |
| Su et al. [18] | "Improved wavelet-based image watermarking through SPIHT" | Watermark bits are embedded in the spatial domain on DC coefficients using 2D-DFT. | **Cover image** Colored Size: 512 x 512 pixels | PSNR = Range [37.6262-38.0535] SSIM= | NCC= Range [0.8413-1.0000] |

| | | | | Watermark Size: 32 x 32 pixels | Range[0.9231-0.9414] | |
| --- | --- | --- | --- | --- | --- | --- |
| Kumar et al. [19] | "Improved wavelet-based image watermarking through SPIHT" | DWT + DCT + SVD Set Partitioning in Hierarchical Tree (SPIHT) and Arnold transform for enhancing robustness. | **Cover image** Size: 512 × 512 pixels **Watermark** Size: 256× 256 pixels | PSNR= Range [25.31- 38.47] SSIM= Range [0.971311-0.999954 ] | NCC= Range [0.9422-0.9990 ] **Attacks Considered** Salt & Pepper, Gaussian Noise, JPEG compression, cropping, Rotation, Scaling attacks, Sharpening Mask, Median Filter, Histogram |
| Kahlessenane Fares et al. [20] | "A robust blind color image watermarking based on Fourier transform domain" | LSB substitution method is used. Watermark is embedded within the mid-band coefficients and frequency components. | **Cover Image** Colored Size: 512 x 512 Pixels **Watermark** Size: random sequence of 45,000 bits | PSNR > 40 dB | NCC= Range[0.6437-0.98265] **Attacks considered** Gaussian noise, Gaussian filter, Histogram, JPEG, rescaling, Rotation, cropping, Gaussian noise, Image sharpening, Blurring, rotation, JPEG compression, salt & Pepper |

| | | | | | noise, Median filter, cropping, Rotation |
|---|---|---|---|---|---|
| Yuan et al. [21] | "New image blind watermarking method based on two-dimensional discrete cosine transform" | Blind watermarking method: Two dimensional DCT is performed on selected blocks. Middle frequency coefficients are embedded. | **Cover Image** Colored Size: 512 × 512 Pixels **Two Colored watermarks** Size: 32 × 32 Pixels | Watermark1 PSNR = range[36.3189-38.2472] SSIM= range [0.9149 - 0.9441] Watermark2 PSNR = Range [36.1885 – 38.4066 ] SSIM = Range [0.9154 – 0.9724 ] | NCC = Range [0.9997 – 1] **Attacks considered :** JPEG(40), JPEG 2000, Gaussian white Noise, salt & pepper Noise, Butter worth low-pass filtering, Median filtering, cropping |
| Anand and Singh [22] | "An improved DWT-SVD domain watermarking for medical information Security" | Medical image watermarking in DWT-SVD domain. Hamming code is put to text watermark before embedding. Chaotic-LZW and hyper Chaotic -LZW are used. | **Cover image** MRI image of Size: 512 x 512 Pixels **Watermark** Size: 256 x 256 Pixels | PSNR = Range [32.6229 - 34.0455] SSIM= Range [0.9950 - 0.9955] | NCC= [0.9724 - 0.9873] **Attacks considered :** Salt & Pepper noise, Gaussian noise, Rotation, JPEG compression, Speckle noise, Cropping, Median filter, Histogram equalization |
| Mishra et al. [23] | "Bi-directional extreme learning | DWT-BELM approach DWT: 4-level | **Cover image** JPEG compressed | PSNR= Range[42.49-42.67] dB | NCC= Range[0.95-1.00] |
| | machine for semi-blind watermarking of compressed images" | Bi-directional ELM (B-ELM): Variant of ELM is modeled with LL4 coefficients for Semi-blind recovery. | image Size: 512 x 512 Pixels<br><br>**Watermark** Binary Size: 32 x 32 Pixels | SSIM= Range [0.9960 - 0.9969] | BER= Range [0.0938 - 0.1074 ]<br><br>**Attacks considered :**<br><br>Low pass Gaussian filter, Median filter, Gaussian noise, Salt & Pepper noise, Rotation, Scaling, Cropping |
| Ambadekar et al. [24] | "Digital Image Watermarking Through Encryption and DWT for Copyright Protection" | DWT and encryption-based watermarking. | **Cover Image** Size: 228 × 228 Pixels<br><br>**Watermark** Size: grayscale image 90 × 90 pixels | PSNR= 54.96 dB | NCC= 0.9749<br><br>**Attacks Considered :**<br><br>Noise, Geometric, Compression |
| Su et al. [30] | "An Approximate Schur Decomposition-based Spatial Domain Color Image Watermarking Method" | Spatial domain watermarking using Schur Decomposition Maximum eigenvalues are approximated and used for embedding and blind extraction. | **Cover Image** Size: 512 x 512 Pixels<br><br>**Watermark** Size: 32 x 32 Pixels | PSNR = Range[40.0691-40.6450] dB<br><br>SSIM = [0.9594-0.9681] | NCC = Range[0.9912-1.0]<br><br>**Attacks Considered :** Rotation, JPEG (30), JPEG 2000, Salt & Peppers noise, Gaussian Noise, Median filtering, Butterworth lowpass filtering, |
| | | | | | Sharpening, Blurring, Scaling, Cropping |
| Rajpal et al. [31] | "Multiple scaling factors based Semi-Blind Watermarking of Grayscale Images using OS-ELM Neural Network" | 4-level DWT decomposition is used. LL4 coefficients are embedded with multiple scaling factors. OS-ELM is used for semi-blind watermarking. | **Cover Image** Size: 512 x 512 Pixels **Watermark** Size: 32 x 32 Pixels | PSNR = [42-43] dB | BER <=0.0029 **Attacks Considered :** JPEG Compression, Rotation, Gaussian noise, Salt & Pepper |
| Hosny et al. [32] | "Parallel Multi-Core CPU and GPU for Fast and Robust Medical Image Watermarking" | Moments of the polar complex exponential transform obtained using Simplified exact kernels are used for the restoration of watermark. | **Cover Image** Size: Colour images and grayscale medical images 256 x 256 Pixels **Watermark** Size: 32 x 32 Pixels | PSNR= Range [40.597 – 53.64 ] dB SSIM= Range [0.933 – 0.980] | NCC= For colored images Range [0.9100 – 1.0] For Grayscale images Range [0.9358 – 1.0] BER= For colored images Range [0 – 0.0156] For Grayscale images Range [0 – 0.0176] **Attacks Considered :** Rotation at various angles, Scaling factor, Translation, Scaling + Rotation, |

| | | | | | Scaling + JPEG compression, Rotation + JPEG compression, JPEG compression, Salt & Pepper Noise, Gaussian Noise, Gaussian filtering, Median filtering |
|---|---|---|---|---|---|
| Taha et al. [33] | "Adaptive Image Watermarking Algorithm Based on an Efficient Perceptual Mapping Model" | Integer-based lifting wavelet transform is used. | **Cover Image Size:** 512 x 512 Pixels  **Watermark Size:** Binary 32 x 32 Pixels | PSNR= Range [32.3236 – 40.5670] dB  SSIM= Range [0.9522 – 0.9883] | NCC= Range [0.7798 – 1]  BER= [0 – 0.1914]  **Attacks Considered :** JPEG compression, Salt & Pepper, Gaussian Noise, Sharpen, Median Filter, Average filter, Rotate, Scale Down, Crop |
| Rajpal et al. [34] | "Fast Digital Watermarking of Uncompressed Colored Images using Bidirectional Extreme | Generation of watermark sequence using DCT using host image  Informed watermarking using Extreme Learning | **Cover Image** Colored Size: 256 x 256 Pixels | PSNR = [39.70- 45.72]  SSIM = [0.9965- 0.9991] | NCC = [0.71- 0.99]  **Attacks Considered :**  Low pass |

| | Learning Machine" | Machine (ELM) and Bi-directional ELM (B-ELM) | **Watermark Size:** 1024 x 1024 Pixels | | Filter, Gaussian Noise, Scaling, Crop, JPEG compression |
|---|---|---|---|---|---|
| Alshoura et al. [35] | "A New Chaotic Image Watermarking Scheme Based on SVD and IWT" | IWT-SVD Hybrid is used. Random key is generated using cover image and the watermark image. This key is utilized for embedding of watermark using chaotic multiple scaling factors (CMSF). | **Cover image** Size : 512 x 512 Pixels  **Watermark** Grayscale image Size: 256 x 256 Pixels | PSNR= Range [46.85 – 52.33] dB | NCC= Range [0.99101-0.99513]  **Attacks Considered :** Cropping, Cutting, Translating, Shifting, Rotating, Scaling, Median filter, Gamma Correction, Median filter, Wiener filter, Histogram equalization, Salt Peppers Noise, Speckle Noise, Gaussian Filter, JPEG Compression |

In Table 1, a study and evaluation of work done in digital watermarking techniques in past are enumerated. The spatial domain and frequency domain techniques are some popular techniques examined in the past. Also, it has been observed that the spatial domain digital watermarking technique is less robust and hence less preferred. The performance of the watermarked image is evaluated through robustness, imperceptibility, security, and capacity. Among these the most preferred criteria, were the visual imperceptibility of the watermarked image and the robustness of the watermarking. In fact, the future work holds scope by combining techniques and using them in hybrid form to not only enhances the robustness of the watermarked image, but it may also reduce the drawbacks of each method considered separately.

## 6. Conclusions

This paper gives an overview of techniques of digital image watermarking along with the detailed classification and characteristics. The various application areas such as medical, remote education, military, electronic voting systems have been presented. It has been seen that data security has become a top priority due to the extensive transmission of digital data. So, for providing authorized data or safeguarding important data, digital watermarking is used. The performance of the watermarked images is evaluated through robustness, imperceptibility, security, and capacity. These are analyzed using peak signal to noise ratio and bit-error ratio. It was observed that robustness was a preferred criterion. Invisible watermarking is carried out for content authentication and proof of ownership. Research groups have preferred frequency domain techniques and have tried to work on balancing between robustness and visual imperceptibility. Through the paper, we analyze various watermarking methods in digital images used in the recent past.